\documentclass[%
 reprint,
 amsmath,amssymb,
 aps,
]{revtex4-1}

\usepackage{graphicx}% Include figure files
\usepackage{dcolumn}% Align table columns on decimal point
\usepackage{bm}% bold math
\usepackage{xcolor}
\usepackage{gensymb}
\begin{document}

\preprint{APS/123-QED}

\title{Analysis of a Waveguide-Fed Metasurface Antenna}% Force line breaks with \\

\author{David R. Smith}
\email{drsmith@duke.edu}
\affiliation{%
Department of Electrical and Computer Engineering, Duke University, Durham, North Carolina, 27708
}%

\author{Okan Yurduseven}%
\affiliation{%
Department of Electrical and Computer Engineering, Duke University, Durham, North Carolina, 27708
}%

\author{Laura Pulido Mancera}
 \altaffiliation[Also at ]{Kymeta Corporation, 12277 134th Court NE, Redmond, Washington 98052, USA.}
\affiliation{%
Department of Electrical and Computer Engineering, Duke University, Durham, North Carolina, 27708
}%

\author{Patrick Bowen}
\altaffiliation[Also at ]{Kymeta Corporation, 12277 134th Court NE, Redmond, Washington 98052, USA.}
\affiliation{%
Department of Electrical and Computer Engineering, Duke University, Durham, North Carolina, 27708
}%

\author{Nathan B. Kundtz}
\affiliation{%
Kymeta Corporation, 12277 134th Court NE, Redmond, Washington 98052, USA.}%

\date{\today}% It is always \today, today,

\begin{abstract}
The metasurface concept has emerged as an advantageous reconfigurable antenna architecture for beam forming and wavefront shaping, with applications that include satellite and terrestrial communications, radar, imaging, and wireless power transfer. The metasurface antenna consists of an array of metamaterial elements distributed over an electrically large structure, each subwavelength in dimension and with subwavelength separation between elements. In the antenna configuration we consider here, the metasurface is excited by the fields from an attached waveguide. Each metamaterial element can be modeled as a polarizable dipole that couples the waveguide mode to radiation modes. Distinct from the phased array and electronically scanned antenna (ESA) architectures, a dynamic metasurface antenna does not require active phase shifters and amplifiers, but rather achieves reconfigurability by shifting the resonance frequency of each individual metamaterial element. Here we derive the basic properties of a one-dimensional waveguide-fed metasurface antenna in the approximation that the metamaterial elements do not perturb the waveguide mode and are non-interacting. We derive analytical approximations for the array factors of the 1D antenna, including the effective polarizabilities needed for amplitude-only, phase-only, and binary constraints. Using full-wave numerical simulations, we confirm the analysis, modeling waveguides with slots or complementary metamaterial elements patterned into one of the surfaces. 
\end{abstract}

\pacs{Valid PACS appear here}% PACS, the Physics and Astronomy
\maketitle

%\tableofcontents

\section{\label{sec:level1}Introduction}
The waveguide-fed metasurface is an emerging concept for aperture antenna design that leverages resonant, subwavelength, radiating elements to generate desired radiation patterns for applications including beam forming for satellite communications  \cite{kundtz_microwave_j_2014, guerci_microwave_j_2014,johnson_ieee_ant_prop_2015,johnson_app_opt_2014}, radio frequency (RF) imaging \cite{hunt2013metamaterial,yurduseven_opt_exp_2016,gollub2017large}, wireless power transfer \cite{smith2017analysis,yurduseven_ieee_access_2017} and synthetic aperture imaging \cite{boyarsky_josaa_2017,watts2017x,pulido_josab_2016}. The use of subwavelength scattering or radiating elements over an aperture enables the effective electric and magnetic current distributions to be conceptualized as continuous, motivating a holographic design approach for the antenna as opposed to the discrete mathematics that would characterize phased arrays and electronically scanned antennas (ESAs) \cite{balanis2016antenna,hansen2009phased,williams1981electronically,fenn2000development}.

One metasurface antenna implementation, introduced by Fong et al. \cite{fong_ieee_ant_prop_2010}, consisted of a structured surface with a spatially varying, artificial impedance. An RF source---a monopole antenna, for example---launches a surface wave onto the surface that has been patterned with the impedance distribution needed to convert the source wave into the desired radiation pattern. The initial impedance distribution is obtained by standard holographic interference techniques, and realized through the use of structured, metamaterial elements. Since the elements are subwavelength in dimension, they can be used to approximate continuous hologram solutions, such as that considered early on by Oliner and Hessel \cite{oliner_ire_trans_ant_prop_1959}. The surface-wave metasurface antenna has proven to be an attractive platform for electrically large and conformal apertures \cite{maci2011metasurfing}, with many antenna variations now demonstrated \cite{minatti_ieee_ant_prop_2011,pandi2015design,minatti_ieee_ant_prop_2015,sun2012gradient,minatti_ieee_ant_prop_2016, patel2011printed}. The concept of slot arrays has also gained traction as a means of enabling beam synthesis for different applications \cite{ando1985radial,liao2012synthesis}. The slot array concept is conventionally based on an array topology, where the element periodicity (or spacing) is comparable with the free-space wavelength $\lambda_0$, typically on the order of $\lambda_0$/2 as opposed to smaller, sub-wavelength element spacing in metasurface antennas.

The metasurface antenna presents an alternative architecture as compared with that of the phased array or ESA. In typical array antennas, radiating antenna modules tile the aperture with roughly a half-wavelength spacing, with control over the phase introduced by active phase shift circuits at each module. By contrast, the metasurface architecture is passive, deriving the phase shift needed for beam steering from the sampled reference wave---a surface wave or waveguide mode, for example, which propagates over the aperture, presenting an advancing phase to each radiating metamaterial element. The metasurface antenna compensates for the loss of independent control over the phase by sampling the aperture with a spatial frequency significantly higher than half the free-space wavelength. In addition, if the metamaterial elements used to form the metasurface array are resonant, some phase delay is added to the radiated wave that can be controlled by tuning the resonances. This latter mechanism of controlling phase has been leveraged in the development of Huygen's metasurfaces \cite{pfeiffer2013metamaterial,wong2014design,epstein_josab_2013} and metasurface holograms \cite{lipworth_opt_exp_2016,zheng2015metasurface,genevet_rep_prog_phys_2015}. Even in cases where little or no additional phase shift is obtained from the elements, the metasurface aperture can nevertheless achieve high-quality beam forming and other wavefront shaping functionality by sampling the phase of the reference wave, often rivaling the performance of more advanced active antenna systems \cite{kundtz_microwave_j_2014, maci2011metasurfing}. 

Our goal here is to present an analysis of the beam forming operation of a waveguide-fed metasurface antenna, under a set of assumptions that enable relatively simple expressions to be found for the key antenna features. We aim to provide a clear path connecting the physics-based polarizibility framework to various antenna metrics, with a focus of achieving closed-form analytical results that illustrate immediately the dependencies of these metrics on the antenna parameters. With this analysis, the waveguide-fed metasurface antenna can be quickly understood and compared with other types of aperture antennas, before more extensive calculations or numerical simulations are pursued. It should be emphasized that our purpose in this paper is not to report a new or improved antenna design but rather present the polarizible dipole framework to analyze waveguide-fed metasurface antennas. The presented technique differs from previous methodologies used to design metasurface antennas in that we make use of a polarizable particle based approach rather than the conventional modulated surface impedance technique. As an example, \cite{oliner_ire_trans_ant_prop_1959} concerns the study of guided waves on a sinusoidally modulated reactive surface, and is a rigorous, self-consistent, quasi-analytic solution to a specific problem. The surface impedance is a continuous function in \cite{oliner_ire_trans_ant_prop_1959}, so that there is no reduction to implementation in this work. That reduction to realizable structures occurred later \cite{fong_ieee_ant_prop_2010, maci2011metasurfing, minatti_ieee_ant_prop_2011} in the context of launching a surface mode that would then radiate as a collimated beam. This is an inherently different type of analysis and structure, and one that depending on a discrete surface impedance that would closely approximate a smooth, continuous function. The guided-wave metasurface provides for arbitrary wave forms, and (as will be shown in this work) can achieve beam forming with discontinuously varying elements (such as the on/off configuration). Consequently, both the method and the structure, in our opinion, are quite distinct from the analytical theory presented in \cite{oliner_ire_trans_ant_prop_1959} and the later physical implementations. Similarly, in \cite{patel2011printed}, the resulting structure is a waveguide-fed, discrete realization of the continuous impedance surface in \cite{oliner_ire_trans_ant_prop_1959} realized using a series of slots with varying slot width. In the dipole language presented in this work, this would be a series of dipoles with varying amplitude, but non-resonant so that there would be no phase variation. That structure is more constrained in scope than the structures analyzed in our present manuscript, where the dipole framework allows us considerable freedom in achieving a wider range of phases and magnitudes, with arbitrary variation.

The geometry we consider, shown in Fig. \ref{fig:metasurface-antenna} below, consists of a one-dimensional waveguide (i.e., propagation allowed in only one direction) that feeds a linear array of radiating, metamaterial elements. Each element is assumed to have a resonance frequency that can be adjusted by varying either the geometry of the element or the local dielectric environment.

The key assumptions made in our model are (1) that the waveguide mode is unperturbed by the elements, and (2) that the elements act as simple, polarizable dipoles, and do not interact with each other. While overly restrictive and generally unrealistic, we demonstrate that the analytical expressions obtained under these assumptions nevertheless are accurate in comparison with full-wave numerical simulations of metasurface antenna implementations. While such agreement is unlikely to persist over all metasurface designs, the results obtained indicate that the analysis presented here provides a useful first pass at a metasurface design, and can be used to build intuition during the design process. 

In Sec. \ref{sec:metasurface-antenna-intro} we introduce the underlying structure of the metasurface antenna and the analysis framework, in which each metamaterial element is conceptually replaced by a polarizable dipole. This modeling approach has been used for waveguide-fed metasurface antennas presented in prior work, and has successfully been implemented in numerical tools for characterizing metasurface apertures \cite{lipworth_app_opt_2015,pulido_awpl_2016}. From this simple model we obtain the radiated far-field pattern. In Sec. \ref{sec:beam-forming}, we extract an array factor from the expression for the far-field, using it to analyze the cases of amplitude-only or Lorentzian-constrained holograms.

Having obtained analytical expressions for the field patterns and presenting several examples of beam forming in Sec. \ref{sec:beam-forming}, we perform full-wave numerical simulations on a slotted waveguide metasurface antenna and a waveguide-fed complementary electric resonator (cELC) metasurface antenna in Sec. \ref{sec:num_sims}. For the specific choices of waveguide and metamaterial elements, we find close agreement between the analytical formulas and the numerical simulations. This agreement implies that the dipole model for the metamaterial elements is valid, and that interactions among the elements are not significant for the structures simulated. For cases where these element-to-element interactions are not negligible, a self-consistent interacting dipole model can be applied for improved accuracy \cite{pulido_awpl_2016}.

\begin{figure}[htbp]
\centering
\mbox{\includegraphics[width=\linewidth]{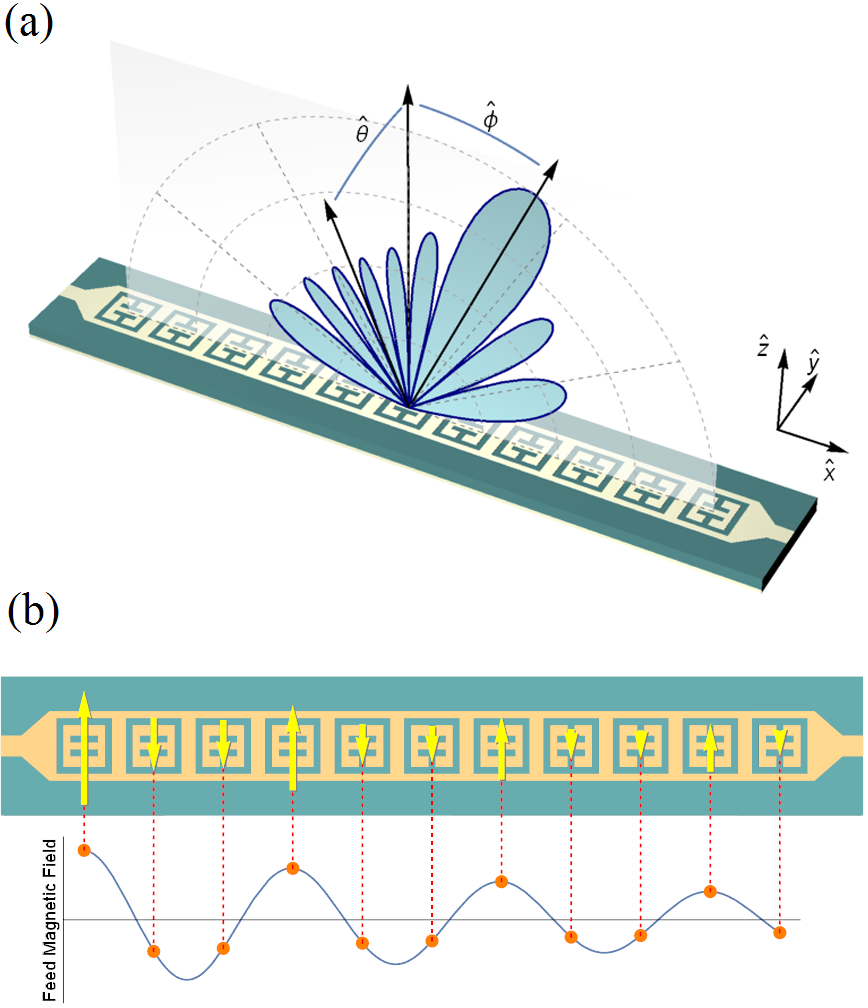}}
\caption{Metasurface antenna (a) depiction of the antenna (b) illustration of the excitation of metamaterial elements by the feed wave.}
\label{fig:metasurface-antenna}
\end{figure}

\section{\label{sec:metasurface-antenna-intro}Metasurface Antenna: Basic Operation}

While the guided wave metasurface antenna can take many different forms, a fairly generic example of the antenna is illustrated in Fig. \ref{fig:metasurface-antenna}(a). In this example, a microstrip transmission line serves to excite an array of complementary metamaterial elements patterned into the upper conductor. The elements shown---complementary electric resonators, or cELCs---have dimensions and spacing significantly smaller than both the free space wavelength $\lambda_0$ and the guided wavelength $\lambda_g$. Complementary metamaterial elements consist of patterns of voids in conducting sheets, forming the Babinet equivalents of metamaterials \cite{falcone_prl_2004,hand_apl_2008,afrooz_motl_2012}. Complementary metamaterial elements patterned into waveguides influence the properties of the bound waveguide mode, but also allow energy to leak out of the waveguide and couple to radiative modes \cite{landy_phot_nano_2013}. For the development of radiating structures, complementary metamaterial elements with effective magnetic response are of interest, since such elements will provide a better radiation efficiency; elements with electric response radiate poorly when embedded in a conducting plane. Numerous types of complementary metamaterials with magnetic response are available, including cELCs and slots \cite{landy_phot_nano_2013}, complementary meander lines \cite{lipworth_opt_exp_2016} and many others. Even more traditional antenna elements, such as iris-fed patch antennas \cite{lee_proc_ieee_2012} can be viewed as magnetic metamaterial elements.

We assume that each of the metamaterial elements is small compared with the free-space wavelength, $\lambda_0$, such that its radiation pattern can be well-approximated by the field radiated from a magnetic dipole \cite{jackson_classical_1999}. The cELCs shown in Fig. \ref{fig:metasurface-antenna}(a), for example, are effectively resonant circuits that produce strong in-plane currents near resonance, which give rise to an effective magnetic surface current with a dominant dipolar response. The metasurface antenna, then, can be modeled as a collection of polarizable point dipoles, each with a frequency-dependent, effective magnetic polarizability $\alpha_m(\omega)$. The dipole at position $x_i$ along the waveguide is assumed to be excited by the magnetic component of the waveguide field at the same point, as illustrated in Fig. \ref{fig:metasurface-antenna}(b).

Each metamaterial element, being essentially a resonant electrical circuit that scatters primarily as a dipole \cite{pulido2017polarizability}, has a polarizability described by the Lorentzian form \cite{Lipworth:13}

\begin{equation}
\alpha_m(\omega) = \frac{F\omega^2}{\omega_0^2-\omega^2+j\Gamma \omega}.
\label{eq:lorentzian}
\end{equation}

\noindent where $F$ is the oscillator strength (real number), $\omega_0$ is the resonance frequency and $\Gamma$ is the damping factor. The position of the dipole can be taken as the center of the cELC or other metamaterial element within the dipole approximation \cite{pulido2017polarizability}. $\omega_0$ relates to the inductance and capacitance of the resonant circuit in the usual manner ($\omega_0=1 / \sqrt{LC}$), and can be tuned by modifying the geometry of the metamaterial element, modifying the local dielectric environment, or by integrating lumped passive or active elements into the circuit. If any of these tuning approaches can be varied via external control, then the metasurface antenna can be reconfigured dynamically \cite{brookner2016metamaterial,pavone2017surface,kundtz_microwave_j_2014}.

The polarizability in Eq. \ref{eq:lorentzian} connects the induced magnetic dipole moment on the metamaterial element, $\vec{m}$, with the local magnetic field of the feeding waveguide mode, $\vec{H}_{ref}$, which we refer to here as the reference wave, in keeping with the holographic description of the antenna presented in the introduction. In the absence of interactions between the metamaterial elements, via either radiation or waveguide modes, the relationship between each metamaterial element and the reference wave is \cite{Lipworth:13}

\begin{equation}
\vec{m}_i=\alpha_{m, i}(\omega) \vec{H}(x_i).
\label{eq:polarizability}
\end{equation}

\noindent From Eqs. \ref{eq:lorentzian} and \ref{eq:polarizability} it can be seen that the field radiated from the element has an amplitude and phase determined by the reference wave $\vec{H}_{ref}$ multiplied by the polarizability of the metamaterial element, which also introduces an additional amplitude and phase advance to the incident reference wave. The phase and amplitude introduced by a metamaterial element are not independent, however, and are intrinsically linked by the inherent Lorentzian resonance in Eq. \ref{eq:lorentzian} as

\begin{equation}
|\alpha_{m}|=\frac{F \omega}{\Gamma}| \cos \gamma |.
\label{eq:phase_and_amplitude}
\end{equation}

\noindent In Eq. \ref{eq:phase_and_amplitude}, $\gamma$ denotes the phase advance introduced by a metamaterial element. Eq. \ref{eq:phase_and_amplitude} is derived directly from Eq. \ref{eq:lorentzian} and can be considered as the defining design equation for the metasurface antenna. Specifically, rather than exercising independent control over the phase and amplitude of the radiated wave at each location over an aperture, the metasurface antenna operates within the constraint imposed by Eq. \ref{eq:phase_and_amplitude}, which effectively limits the range of phase values that can be added to the reference wave. For example, Eq. \ref{eq:phase_and_amplitude} shows that the phase advance is limited to $\pm 90^\circ$, which is half the full $360^\circ$ range normally required for efficient holograms and diffractive optical elements; moreover, for phase advances near $\pm 90^\circ$, the amplitude approaches zero, suggesting the usable range of phase advances is actually much smaller than $180^\circ$.  While this constraint may at first appear severe, the phase advance of the reference wave combined with sub-wavelength sampling of the aperture can compensate considerably for the loss of independent control over the phase and amplitude of each radiating element. The sampling of the reference wave at each metamaterial element location is depicted in Fig. \ref{fig:metasurface-antenna}(b). Since the resonance of a metamaterial element can be dynamically tuned by numerous modalities, including voltage-controlled semiconductor components such as diodes, transistors and varactors, a metasurface antenna can be reconfigurable with extremely low power requirements and without the need for complex active circuitry.

Each of the metamaterial elements radiates as a magnetic dipole, with a far-field pattern given by

\begin{equation}
\vec{H}_{rad}=-\frac{\omega^2 m}{4 \pi |\vec{r}-\vec{r'}|}\cos \theta  e^{-j k|\vec{r}-\vec{r'}|+j \omega t} \hat{\theta}.
\label{eq:element_radiation_pattern}
\end{equation}

\noindent We have chosen $\hat{\theta}$ such that $\cos \theta=1$ in the broadside direction. $\vec{r}'$ is a vector that locates the position of the source. Within the microstrip or waveguide, the transverse component of the magnetic field at the position of a given metamaterial element is the predominant excitation, and (in the absence of losses or reflection) will have a sinusoidal dependence as a function of distance along the waveguide. For the initial analysis, we assume the reference wave has the dependence \cite{balanis2016antenna}.

\begin{equation}
\vec{H}_{ref}=H_0 e^{-j \beta x}\hat{y},
\label{eq:ref_wave_1}
\end{equation}

\noindent and is not perturbed by the scattering from the cELC elements. $\beta$ is the propagation constant for the waveguide mode, and can be written as $n_{g}\omega/c$, where $n_{g}$ is the waveguide index.

The metamaterial elements can be thought of as sampling the reference wave and aperture at locations designated by $x_i=i d$, where $i$ is an integer and $d$ is the spacing between any two adjacent elements. Then, the far-field radiation pattern from the metasurface antenna can be approximated by superposing the fields sourced by all of the elements:

\begin{equation}
\vec{H}_{rad}=-H_0 \frac{\omega^2}{4 \pi r}\cos \theta \sum_{i=1}^{N} \alpha_{m,i}(\omega)e^{-j \beta x_i}e^{-jk|\vec{r}-x_i \hat{x}|}\hat{\theta},
\label{eq:radiated_field}
\end{equation}

In Eq. \ref{eq:radiated_field} the $|\vec{r}-\vec{r}'|\approx r$ approximation was used in the Green's function denominator. \noindent Eq. \ref{eq:radiated_field} can be used to obtain an approximate field pattern for the one-dimensional metasurface antenna. The applicability of Eq. \ref{eq:radiated_field} depends on a number of factors. In particular, for an actual implementation, the effective polarizabilities $\alpha_{m,i}$ for the metamaterial elements must be determined accurately, which requires an extracting method from either measurements or numerical simulations. Such methods have been presented elsewhere \cite{scher_meta_2009,karamanos_adv_electromagnetics_2012} for periodic metasurfaces under plane wave incidence, and in \cite{pulido2017polarizability} for waveguide-fed metasurfaces, and provide a path towards a highly accurate modeling platform for metasurface apertures. Since we are more interested here in general trends and behavior, we do not pursue extraction methods further in the present analysis.

Once the polarizability has been assigned, if it is further assumed that the elements do not perturb the waveguide mode and do not interact with each other, then Eq. \ref{eq:radiated_field} will be a good approximation for the radiated field. The assumptions outlined above can be surprisingly useful to obtain a good description of the behavior of a metasurface antenna. Of the three approximations described above, the decay of the reference wave due to the radiation of the elements is the most important and can be taken into account in a number of ways, one of which will be described below. 

\section{Beam Forming}
\label{sec:beam-forming}

The metasurface antenna design approach can be thought of as being somewhere between the discrete sampling of the aperture used in array antennas, and the continuous sampling that would motivate holographic methods. The distinct approach to the metasurface antenna follows because the scale of and spacing ($d$) between the radiating metamaterial elements are significantly smaller than the typical $\lambda_0/2$ spacing associated with phased arrays, though often practically limited to dimensions on the order of $\lambda_0/10<d<\lambda_0/5$ \cite{smith2017analysis}. Because independent and complete control over the phase at each radiating point is not possible, the subwavelength sampling of the aperture is crucial to obtain the best performance of metasurface antennas.

A typical function of a reconfigurable antenna is to produce a collimated beam in the far-field with a desired direction $(\phi_0, \theta_0)$ (i.e., a beam with narrow angular spread). We begin by considering the design---the specific polarizability distribution---needed for the waveguide-fed metasurface antenna to form such a beam. To start the analysis, we make the usual assumption that the points of observation are in the far-field relative to the aperture, so that we can approximate Eq. \ref{eq:radiated_field} as

\begin{equation}
\vec{H}_{rad}=H_0 \frac{\omega^2}{4 \pi r}e^{-jkr} \cos \theta \sum_{i=1}^{N} \alpha_{m,i}(\omega)e^{-j \beta x_i}e^{-j k x_i \sin \phi}\hat{\theta}.
\label{eq:frauenhofer_field}
\end{equation}

\noindent Known as the Fraunhofer approximation \cite{balanis2016antenna}, we have made use of $|\vec{r}-\vec{r}'|\approx r \; \sqrt[]{1-2\vec{r}\cdot \vec{r}'/r^2}$, which allows a separation to be made between the radial and the angular dependences of the far-fields. In particular, the radial dependence becomes a simple prefactor, with the angular distribution of the field determined by a metasurface array factor $AF$ of the form

\begin{equation}
AF(\phi,\theta)= \cos \theta \sum_{i=1}^{N} \alpha_{m,i}(\omega)e^{-j \beta x_i}e^{-j k x_i \sin \phi}.
\label{eq:array_factor_definition}
\end{equation}

\noindent Eq. \ref{eq:array_factor_definition} can be used to calculate far-field radiation patterns for the metasurface antenna. Note that Eq. \ref{eq:array_factor_definition} is identical to the array factor used for array antennas \cite{hansen2009phased}, aside from the greater sampling that characterizes the metasurface antenna.

To form a collimated beam in the direction $\phi_0$, the polarizabilities (or weights) of the metamaterial elements must be chosen such that the waves from each of the radiators are in phase and interfere constructively in the chosen direction. We derive the necessary weights by determining the field distribution needed in the plane of the aperture. A plane wave propagating in the direction $\phi_0$, for example, has the form 

\begin{equation}
\vec{H}_{pw}= H_{pw}e^{-j (k_x x+k_z z)}\hat{\theta}.
\label{eq:plane_wave}
\end{equation}

\noindent In Eq. \ref{eq:plane_wave}, $k_x$ and $k_z$ denote the wavenumbers in the x- and z-axes respectively. In the plane of the antenna ($z=0$), then, the field must have the dependence $\exp(-j k x \sin \phi_0)$, where $\phi_0$ is the angle of propagation with respect to the surface normal of the antenna (broadside direction). Thus, by comparison with Eq. \ref{eq:array_factor_definition}, we see the desired weights required to obtain such a field distribution must be

\begin{equation}
\alpha_{m,i}(\omega)=e^{j \beta x_i}e^{j k x_i \sin \phi_0}.
\label{eq:plane_wave_weights}
\end{equation}

\noindent In this highly idealized approach to determining the polarizabilities, we see that the polarizabilities are chosen to compensate for the propagation of the waveguide mode, then to add the phase and amplitude distribution required to generate the directed beam. Substituting Eq. \ref{eq:plane_wave_weights} into Eq. \ref{eq:array_factor_definition}, we obtain

\begin{equation}
AF(\phi,\theta)= \cos \theta \sum_{i=1}^{N} e^{-j k x_i (\sin \phi-\sin \phi_0)}.
\label{eq:plane_wave_fp_ideal}
\end{equation}

\noindent The array factor of Eq. \ref{eq:plane_wave_fp_ideal} predicts a radiation pattern highly peaked in the $\phi_0$ direction, with a series of side lobes that fall off away from the central peak.

The polarizabilities, or weights, determined by Eq. \ref{eq:plane_wave_weights} would require full control over the phase (with the amplitude being constant) of the transmitted radiation at each position $x_i$ along the metasurface antenna, which is generally not feasible given the constraints of the metamaterial elements expressed by Eq. \ref{eq:phase_and_amplitude}. It is at this point that we  move away from conventional phased array design methodology, and seek alternative weighting functions that will enable the same beam forming capabilities with the metasurface architecture.

\subsection{Amplitude Only Hologram}

In considering applying various weight distributions to the metamaterial elements, it is useful first to recall how the metamaterial elements can be modified. Two possible routes for element tuning are to shift the resonance frequency by changing the capacitance or inductance of the resonator circuit, or to change the damping factor by modifying the resistance of the circuit. Assuming the resonance frequency shift or damping occurs in response to an applied voltage bias $V_B$, then we can write $\omega_0(V_B)$ or $\Gamma(V_B)$. A tuning state, or mask, for the metasurface antenna then corresponds to the resonance frequency and damping factor at each of the metamaterial elements, which will map to polarizabilties through Eq. \ref{eq:lorentzian}. The coupling between the reference wave and the metamaterial element, encapsulated in the factor $F$ in Eq. \ref{eq:lorentzian}, can also be potentially tuned, though such a mechanism will be more involved.

If the metamaterial element is near resonance, then adjusting the damping factor will effectively tune the amplitude of the metamaterial element ($\alpha_m(\omega)\approx -j F \omega /\Gamma$). Similarly, changing the coupling factor $F$ also tunes the amplitude directly, without significant phase shift. Given that such a tuning modality is possible with the metasurface antenna, it is useful to consider the prospect of amplitude-only tuning.

To convert the complex weight function of Eq. \ref{eq:plane_wave_weights} to an amplitude only weight function, we consider taking the real part of Eq. \ref{eq:plane_wave_weights} in the form

\begin{multline}
\alpha_{m,i}(\omega)=X_i+M_i \cos(\beta x_i+k x_i \sin \phi_0) \\ =X_i+M_i(\frac{e^{j \beta x_i}e^{j k x_i \sin \phi_0}}{2}+\frac{e^{-j \beta x_i}e^{-j k x_i \sin \phi_0}}{2}),
\label{eq:plane_wave_weights_amp}
\end{multline}

\noindent where $X_i$ and $M_i$ are real and positive. We allow for both a constant and a modulation term, since the amplitude-only weights must be positive, and there will be some limited tuning range achievable by either modifying the resonance frequency or damping factor. For a practical amplitude-only weight distribution, we assume $X_i \geq M_i$. The array factor, Eq. \ref{eq:array_factor_definition}, then takes the form

\begin{multline}
AF(\phi)=  \sum_{i=1}^{N}X_i e^{-j x_i (k \sin \phi+\beta)} \\ +\sum_{i=1}^{N}\frac{M_i}{2} e^{-j k x_i (\sin \phi-\sin \phi_0)} \\ +\sum_{i=1}^{N}\frac{M_i}{2} e^{-j k x_i \sin \phi} e^{-j k x_i \sin \phi_0} e^{-j 2 \beta x_i}.
\label{eq:plane_wave_fp_amp}
\end{multline}

\noindent From Eq. \ref{eq:plane_wave_fp_amp}, it is clear that the second term, which is identical to the ideal distribution of Eq. \ref{eq:plane_wave_fp_ideal}, produces the desired beam. However, the first and third sums can potentially produce additional, undesired beams, and therefore we must assess the impact of these additional terms \cite{johnson_ieee_ant_prop_2015}.

The formation of a beam occurs for a given sum when the argument of the complex exponential vanishes; when this condition occurs, the fields from all elements interfere constructively \cite{balanis2011modern,schwering1983design}. We can obtain the possible beam directions from Eq. \ref{eq:plane_wave_fp_amp}, then, by noting the angle $\phi$ where the arguments in the exponentials are zero, or

\begin{equation}
\begin{gathered}
\phi = \sin^{-1}(-n_{g}) \\
\phi=\phi_0 \\
\phi=\sin^{-1}(-2 n_{g} - \sin \phi).
\end{gathered}
\label{eq:plane_wave_fp_amp_only_conditions}
\end{equation}

\noindent where $n_g=\beta/k$ is the refractive index inside the waveguide. From Eq. \ref{eq:plane_wave_fp_amp_only_conditions}, it can be seen that while two unwanted beams are possible, both can be suppressed if the waveguide index is large enough, since the absolute value of the argument of the $\sin^{-1}$ will be greater than unity. The first and third terms of Eq. \ref{eq:plane_wave_fp_amp_only_conditions} never produce a beam, since $n_{g} \geq 1$. Note that even though a secondary beam is not formed from the additional terms in Eq. \ref{eq:plane_wave_fp_amp_only_conditions}, there is nevertheless a coherence condition. We refer to that condition as producing a nonpropagating mode, since the propagation vector corresponding to this condition will be evanescent. An alternative but equivalent description of this condition is that the beam has moved to \textit{invisible space}, using language common in array antenna theory \cite{zinka_ieee_ant_prop_2010}.

\begin{figure}[htbp]
\centering
\mbox{\includegraphics[width=\linewidth]{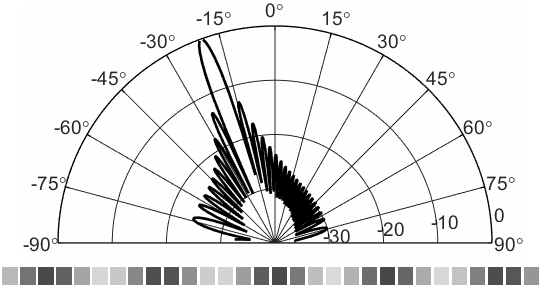}}
\caption{Illustration of beam steering by a metasurface antenna, assuming amplitude-only control over the polarizabilities. Logarithmic scale (dB). The parameters for the calculation are shown in Table \ref{tab:parameters_amp_only_beam}. The bar chart below the plot illustrates the magnitude of the weights for the first 30 elements.}
\label{fig:beam_steer_amp_only}
\end{figure}

We illustrate the beam forming capability of the metasurface antenna with an amplitude-only distribution of weights, as in Eq. \ref{eq:plane_wave_weights_amp} ($X_i=1$, $M_i=0.5$), in Fig. \ref{fig:beam_steer_amp_only}. For this calculation, the operating frequency was assumed to be 10 GHz, with metamaterial element spacing assumed to be 3 mm ($\lambda_0/10$). The waveguide index was chosen as $n_g=2.5$, with the rest of the parameters summarized as in Table \ref{tab:parameters_amp_only_beam}. The value for the waveguide index is somewhat impractically large, and was chosen arbitrarily so that we can illustrate beam forming here without concern of secondary beams being excited. Such a large value for the waveguide index is excessive, however, and quite good beam forming performance can be achieved with the moderate values of waveguide index that are typical.  

As Fig. \ref{fig:beam_steer_amp_only} shows, the beam formed is what would typically be expected from an array antenna, though only the amplitude of each metamaterial element is varied. The chart at the bottom of the plot depicts schematically the relative amplitudes, $\alpha_{m,i}$, for the first 30 elements. From Eq. \ref{eq:plane_wave_fp_amp}, the amplitude distribution is periodic, with $\max(\alpha_{m,i})=1.5$ and $\min(\alpha_{m,i})=0.5$. Such a weight distribution has a close connection with a blazed grating, which diffracts an incident beam while suppressing higher order beams \cite{schwering1983design}.

\begin{table}[htbp]
\centering
\caption{\bf Parameters for the Amplitude Only Metasurface Antenna}
\begin{tabular}{ccc}
\hline
parameter & value & units \\
\hline
operating frequency & 10 & GHz \\
cell size & 3 & mm \\
number of cells & 160 & - \\
operating wavelength & 3 & cm \\
guide index ($n_g$) & 2.5 & - \\
aperture size & 48 & cm \\
min amplitude & 0.5 & - \\
max amplitude & 1.5 & - \\
\hline
\end{tabular}
  \label{tab:parameters_amp_only_beam}
\end{table}

The beam profile shown in Fig. \ref{fig:beam_steer_amp_only} is representative for nearly all scan angles (except those at extreme angles near $\pm 90^{\circ}$, with the beam width increasing for larger scan angles away from the broadside direction due to the $\cos \phi$ factor related to aperture loss \cite{smith2017analysis}). The non-propagating terms in Eq. \ref{eq:plane_wave_fp_amp_only_conditions} do not significantly impair the characteristics of the main beam, but lead to increased side lobe levels as the ratio $\max(\alpha_{m,i}) / \min(\alpha_{m,i})$ tends towards unity (not shown here). Analyzing Fig. \ref{fig:beam_steer_amp_only}, the half-power-beam-width (HPBW) of the amplitude only hologram metasurface antenna is measured to be 3.67\degree while the first sidelobe level is -12.94 dB with the beam pointing at $\phi_0$=-20\degree.

If we ignore the sums corresponding to the non-propagating modes in the array factor of Eq. \ref{eq:plane_wave_fp_amp}, then we can sum the remaining series in the usual way using

\begin{equation}
\sum_{i=1}^{N} e^{-j k x_i (\sin \phi-\sin \phi_0)}=\left(\frac{u^N-1}{u-1}\right),
\label{eq:sum_definition}
\end{equation}

\noindent where $u=\exp [-j k d(\sin \phi-\sin \phi_0)]$. Thus, the intensity of the far-field radiation pattern has the usual form

\begin{equation}
AF(\phi)=\frac{\sin^2[\frac{N}{2}kd(\sin \phi-\sin \phi_0)]}{\sin^2[\frac{1}{2}kd(\sin \phi-\sin \phi_0)]}.
\label{eq:plane_wave_fp_amp_analytic}
\end{equation}

\noindent A plot of the far-field radiation pattern using Eq. \ref{eq:plane_wave_fp_amp_analytic} (not shown) provides an identical result as in Fig. \ref{fig:beam_steer_amp_only}.

Using Eq. \ref{eq:plane_wave_fp_amp_analytic}, it is possible to derive the usual approximations for the beam width and side lobe levels ($\Delta \phi =2.782 \lambda/(2 \pi/L)$ and -13.45 dB, respectively, where $L$ is the total size of the aperture) \cite{balanis2016antenna}. Examining Fig. \ref{fig:beam_steer_amp_only}, it can be seen these values are relatively accurate for the metasurface aperture. To have a slightly more accurate expression for the beam width, we can assume the beam is tightly directed around $\phi_0$, such that $\phi=\phi_0+\Delta \phi$ \cite{goodman2005introduction}. Then, assuming we can make the approximation $\sin \Delta \phi \approx \Delta \phi$, we have $\sin \phi - \sin \phi_0 \approx \cos \phi_0 \Delta \phi$, and Eq. \ref{eq:plane_wave_fp_amp_analytic} becomes

\begin{equation}
AF(\phi)=\frac{\sin^2[\frac{N}{2}kd\cos \phi_0 \Delta \phi]}{\sin^2[\frac{1}{2}kd\cos \phi_0 \Delta \phi]}.
\label{eq:plane_wave_fp_amp_analytic_ang}
\end{equation}

\noindent From this equation, we see that the effective aperture is reduced by the familiar $\cos \phi_0$ factor, and the beam width expands accordingly as $\Delta \phi =2.782 \lambda/(2 \pi/L'$), where $L'=L \cos \phi_0$.

\subsection{Binary Amplitude Hologram}

One of the more easily metamaterial tuning methods is to toggle each metamaterial element between an ``on" state and an ``off" state, such that there are only two possible amplitudes for each element {\cite{yurduseven_ieee_access_2017}. This binary distribution can be achieved, for example, by switching the resonance frequency of an element between that of the operating frequency and another frequency far away from the operating frequency. There are many possible distributions of on/off elements that can potentially provide a directed beam solution, and numerous ways in which to arrive at such distributions, including brute-force optimization. We consider here a simple approach, again motivated by holographic and diffraction optics. We force the continuous amplitude-only distribution to binary by applying a Heaviside step function to Eq. \ref{eq:plane_wave_weights_amp}, or

\begin{equation}
\alpha_{m,i}(\omega)=X_i+M_i \Theta_H[\cos(\beta x_i+k x_i \sin \phi_0)],
\label{eq:plane_wave_weights_amp_bin}
\end{equation}

\noindent where $\Theta_H(x)=0$ if $x<0$ and $\Theta_H(x)=1$ if $x>0$. Eq. \ref{eq:plane_wave_weights_amp_bin} thus represents an offset square wave, which we be can re-expressed by using the Fourier series relationship as

\begin{equation}
\Theta_H[\cos(qx)]=\frac{1}{2}+\frac{4}{\pi}\sum_{m=1,3,5...}^{\infty}\frac{1}{m}\sin (mqx),
\label{eq:square_wave_ft}
\end{equation}

\noindent obtaining

\begin{equation}
\alpha_{m,i}(\omega)=X_i+M_i\sum_{m=1,3,5...}^{\infty}\frac{1}{m}\sin [m(\beta+k \sin \phi_0)x_i],
\label{eq:square_wave_ft}
\end{equation}

\noindent where we have moved the constant terms into the overall constants $X_i$ and $M_i$. Comparing with the derivation that led to Eq. \ref{eq:plane_wave_fp_amp}, we see that there are now an infinite number of terms, each with many potential beams. Given that the binary amplitude distribution is analogous to a non-blazed grating, it is not surprising that the angular spectrum may include one or more diffracted orders. In fact, we can immediately write all of the conditions for which collimated beams are possible as

\begin{equation}
\begin{gathered}
\phi = \sin^{-1}(-n_g) \\
\phi =\phi_0 \\
\phi =\sin^{-1}(-2n_g - \sin \phi_0) \\
\phi =\sin^{-1}[(m-1)n_g +m \sin \phi_0] \\
\phi =\sin^{-1}[-(m+1)n_g -m \sin \phi_0] 
\end{gathered},
\label{eq:plane_wave_fp_amp_conditions}
\end{equation}

\noindent where $m=3,5,7\ldots$. The additional conditions make it somewhat more likely that a diffracted order can appear, but most higher orders continue to be rejected if the waveguide index $n_g$ is large enough. We present the calculation for a beam directed to $\phi_0$=-20\degree, same parameters as in Fig. \ref{fig:beam_steer_amp_only}, but with the weight factors now selected according to Eq. \ref{eq:plane_wave_weights_amp_bin}. The resulting binary weight distribution produces a single beam, but with an increase in the overall side lobe levels. Depending on the application, the side lobe levels associated with this straightforward design may be acceptable, but can certainly be decreased using other optimization methods \cite{johnson_ieee_ant_prop_2015}. Aside from the higher side lobe levels, the result shown in Fig. \ref{fig:beam_steer_bin} is similar to all other scan angles over the entire half plane. Analyzing Fig. \ref{fig:beam_steer_bin}, the HPBW of the binary hologram metasurface antenna is measured as 3.71\degree while the first sidelobe level is -12.75 dB with the beam pointing at $\phi_0$=-20\degree.

\begin{figure}[htbp]
\centering
\mbox{\includegraphics[width=\linewidth]{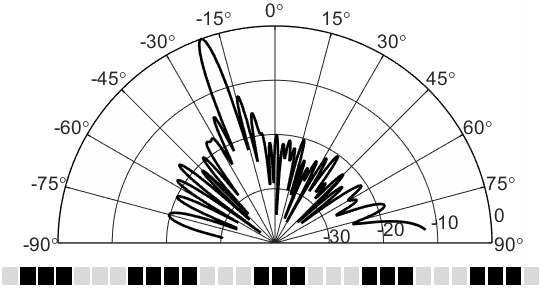}}
\caption{Illustration of beam steering by a metasurface antenna, assuming binary-amplitude control over the polarizabilities. Logarithmic scale (dB). The parameters for the calculation are shown in Table \ref{tab:parameters_amp_only_beam}. The bar chart below the plot illustrates the magnitude of the weights for the first 30 elements.}
\label{fig:beam_steer_bin}
\end{figure}

\subsection{Attenuation of Reference Wave}

In any metasurface antenna, the reference wave will decay along the direction of propagation due to energy loss from both radiative and resistive losses. If we neglect resistive losses as well as the reflection of the reference wave from the metamaterial elements, we can obtain a first approximation to the attenuation length. Since each metamaterial element behaves as a polarizable magnetic dipole, the power dissipated at element $i$ can be written 

\begin{equation}
P_{dis}=-\omega \mu_0 \frac{1}{2} \operatorname{Im}\left\lbrace \vec{m}_i \cdot \vec{H}_{i}^* \right\rbrace=\omega \mu_0 \frac{1}{2} |H_i|^2 \operatorname{Im}\left\lbrace \alpha_{m,i} \right\rbrace,
\label{eq:dipole_dissipated_power}
\end{equation}

\noindent where we have used Eq. \ref{eq:polarizability}. The power per cross-sectional area---or intensity---in the reference wave incident on the element can be found from Poynting's theorem as $I_i=(\vec{E_i}\times\vec{H_i}^*) \cdot \hat{x}$, while the intensity at the subsequent element is $I_{i+1}=(\vec{E_{i+1}}\times\vec{H_{i+1}}^*) \cdot \hat{x}$. Assuming no reflections and that energy lost between the two elements is either radiated or dissipated, We arrive at the relationship

\begin{equation}
|H_i|^2-|H_{i+1}|^2=-\omega \mu_0 \frac{\operatorname{Im}\left\lbrace \alpha_{m,i} \right\rbrace}{2 A_c \operatorname{Re}\left\lbrace \eta \right\rbrace}|H_i|^2,
\label{eq:energy_balance}
\end{equation}

\noindent where $\eta$ is the characteristic impedance of the waveguide and $A_c$ is the cross-sectional area of the guide. If the field varies over the area, then $A_c$ will become a factor that represents an effective area. If the term multiplying $|H_i|^2$ on the right-hand side is sufficiently small, then we can write the following equation, where $d$ denotes the element spacing.

\begin{equation}
\frac{d|H|}{dx}=-\frac{\omega \mu_0}{4 d A_c} \frac{\operatorname{Im}\left\lbrace \alpha_{m} \right\rbrace}{\operatorname{Re}\left\lbrace \eta \right\rbrace}I.
\label{eq:energy_balance_cont}
\end{equation}

\noindent In writing this last equation, we have assumed that the polarizability is the same for each metamaterial element. This will in general not be the case, but presumably it will be possible to arrive at an averaged value for the attenuation coefficient of the magnetic field, which, from Eq. \ref{eq:energy_balance_cont}, is

\begin{equation}
\bar{\alpha}=\frac{\omega \mu_0}{4 d A_c} \frac{\operatorname{Im}\left\lbrace \alpha_{m} \right\rbrace}{\operatorname{Re}\left\lbrace \eta \right\rbrace}.
\label{eq:attenuation_coefficient}
\end{equation}

Having now an expression for the attenuation of the reference wave, we can write the array factor (Eq. \ref{eq:plane_wave_fp_ideal}) as

\begin{equation}
AF(\phi,\theta)= \cos \theta \sum_{i=1}^{N} e^{-\bar{\alpha}x_i}e^{-j k x_i (\sin \phi-\sin \phi_0)}.
\label{eq:plane_wave_fp_attenuation}
\end{equation}

\noindent As before, we can perform the summation, obtaining

\begin{equation}
AF(\phi)= \frac{e^{-N\bar{\alpha}d}e^{-j N k d (\sin \phi-\sin \phi_0)}-1}{e^{-\bar{\alpha}d}e^{-j k d (\sin \phi-\sin \phi_0)}-1}.
\label{eq:plane_wave_fp_attenuation_analytic}
\end{equation}

\noindent It can be expected that, for an electrically large aperture, the field has decayed significantly by the end of the antenna, such that $\exp(-N \bar{\alpha}d)\approx 0$. The beam width and other properties, then, no longer depend on the total aperture size $L=Nd$, but rather on an effective aperture size dictated by the attenuation length $\delta=1/\bar{\alpha}$. Assuming that the value $\bar{\alpha}d<<1$, we arrive at the approximate HPBW as

\begin{equation}
\Delta \phi=\frac{1}{\pi}\frac{\lambda}{\delta}.
\label{eq:beam_width_att}
\end{equation}

The impact of attenuation can be seen in the calculated radiation pattern in Fig. \ref{fig:beam_steer_amp_only_att_ref}, where $\alpha$=6. One immediate feature is the loss of articulated nodes and sidelobes, due to the lack of zeros in the array factor. The beam width is now determined not by the full aperture, but instead through Eq. \ref{eq:beam_width_att}. Analyzing Fig. \ref{fig:beam_steer_amp_only_att_ref}, the HPBW of the metasurface antenna with radiation damping included is measured to be 4.6\degree, wider than the HPBW of the scenarios studied in Fig. \ref{fig:beam_steer_amp_only} and Fig. \ref{fig:beam_steer_bin} where the amplitude decay of the reference wave was not taken into account.

\begin{figure}[htbp]
\centering
\mbox{\includegraphics[width=\linewidth]{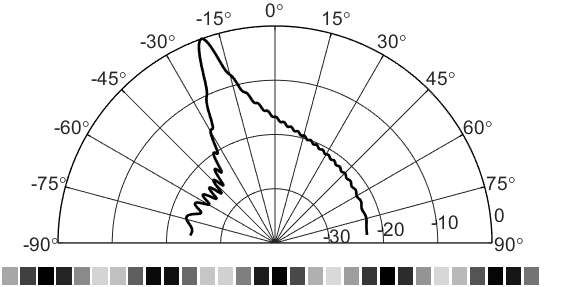}}
\caption{Illustration of beam steering by a metasurface antenna, with radiation damping included. Logarithmic scale (dB). The parameters for the calculation are shown in Table \ref{tab:parameters_amp_only_beam}, except that now an attenuation factor for the waveguide mode has been included. The bar chart below the plot illustrates the magnitude of the weights for the first 30 elements.}
\label{fig:beam_steer_amp_only_att_ref}
\end{figure}

\subsection{Lorentzian Constrained Phase Hologram}

It is well-known in holography and beam forming that control over the phase generally provides a better beam or image quality than what can be accomplished using amplitude control \cite{hariharan1996optical}. For the metasurface antenna, however, the phase and amplitude of the weights are inextricably linked through the Lorentzian resonance of the metamaterial element, which leads to the relationship in Eq. \ref{eq:phase_and_amplitude}. If phase tuning of a metamaterial element is implemented, then the amplitude will necessarily vary as a function of phase. To obtain an efficient hologram, the phase should vary over the range $0\leq \phi \leq 2 \pi$; however, a single Lorentzian resonator is restricted to the range $0 \leq \phi \leq \pi$, meaning that there will inevitably be field distributions not accessible with the Lorentzian constrained metamaterial elements. Still, it is useful to assess here if there are gains that can be achieved by using the phase tuning possible with a resonant metamaterial element.

For beam forming, we seek weights of constant amplitude and linearly increasing phase, or

\begin{equation}
\alpha_{m,i}=e^{j \Psi_i},
\label{eq:ideal_phase_weights}
\end{equation}

\noindent where $\Psi_i=\beta x_i+k x_i \sin \phi_O$. It is instructive to plot Eq. \ref{eq:ideal_phase_weights} as a curve in the complex $\alpha_m$ space, as shown in Fig. \ref{fig:ideal_to_lorentzian_map}; there, it can be seen that the ideal polarizability values lie on a circle with unit radius and centered at the origin. The available range for Lorentzian-constrained polarizabilities, however, (Eq. \ref{eq:phase_and_amplitude}) plotted in the complex plane forms a circle with unity diameter and centered at $\alpha_m=0.5j$. A number of strategies could be followed to map the ideal polarizabilities to a set of constrained polarizabilities; while none of the constrained distributions will lead to perfect beam formation, it may be possible to optimize for certain metrics given the available freedom. 

\begin{figure}[htbp]
\centering
\mbox{\includegraphics[width=\linewidth]{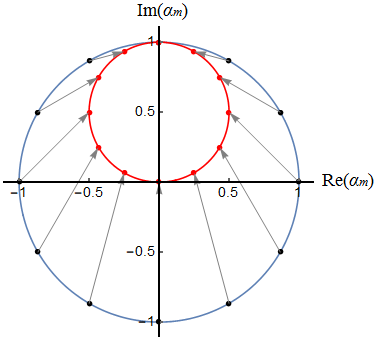}}
\caption{Plot in the complex plane of the ideal weights (outer circle, blue, Eq. \ref{eq:ideal_phase_weights}) and the Lorentzian constrained weights (inner circle, red, Eq. \ref{eq:constrained_phase_weights}). Arrows indicate the mapping between the ideal and constrained points.}
\label{fig:ideal_to_lorentzian_map}
\end{figure}

Rather than pursuing a more extensive optimization, we seek here a simple mapping from the ideal polarizability distribution to the constrained distribution. Consider the weighting function:

\begin{equation}
\alpha_{m,i}=\frac{j+e^{j \Psi_i}}{2},
\label{eq:constrained_phase_weights}
\end{equation}

\noindent it can easily be verified that $0^\circ \leq \angle \alpha_{m,i} \leq 180^\circ$. Moreover, the amplitude of the weights satisfies the constraint of Eq. \ref{eq:phase_and_amplitude}, or

\begin{equation}
|\alpha_{m,i}|=|\cos(\Psi_i /2)|.
\label{eq:constrained_phase_weights_amp}
\end{equation}

Using the Lorentzian-constrained weights of Eq. \ref{eq:constrained_phase_weights}, we arrive at the array factor

\begin{equation}
AF(\phi)=\frac{1}{2} \sum_{i=1}^{N} \left( j e^{j(-\beta x_i-k x_i \sin \phi)}-e^{j k x_i(\sin \phi_0-\sin \phi)} \right).
\label{eq:lorentzian_constrained_af}
\end{equation}

\noindent Eq. \ref{eq:lorentzian_constrained_af} is used to plot the field pattern in Fig. \ref{fig:beam_steer_lorentzian_phase}, which is seen to have good quality despite the amplitude and phase limitations. As with the other scenarios in this section, Fig. \ref{fig:beam_steer_lorentzian_phase} is representative, with the calculated patterns for other steering angles appearing similar.

The array factor, Eq. \ref{eq:lorentzian_constrained_af}, has only one other term that can give rise to an additional beam, so that it can be expected to have reasonably good performance. Additional optimization should potentially improve the situation further. Analyzing Fig. \ref{fig:beam_steer_lorentzian_phase}, the HPBW of the Lorentzian constrained phase hologram metasurface antenna is measured as 3.6\degree while the first sidelobe level is -13.37 dB, suggesting a reduction in the HPBW and sidelobe levels in comparison to the scenarios studied in Fig. \ref{fig:beam_steer_amp_only} and Fig. \ref{fig:beam_steer_bin}. The steered beam of the antenna points at $\phi_0$=-20\degree.

\begin{figure}[htbp]
\centering
\mbox{\includegraphics[width=\linewidth]{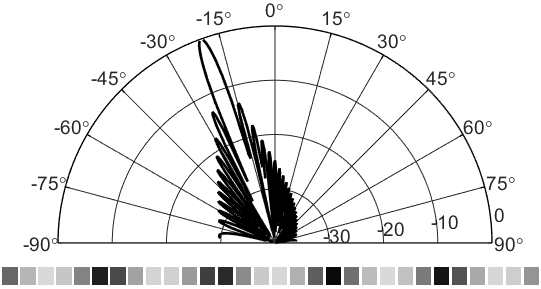}}
\caption{Illustration of beam steering by a metasurface antenna with Lorentzian-constrained phase control over the polarizabilities. Logarithmic scale (dB). The parameters for the calculation are shown in Table \ref{tab:parameters_amp_only_beam}. The bar chart below the plot illustrates the magnitude of the weights for the first 30 elements.}
\label{fig:beam_steer_lorentzian_phase}
\end{figure}

\section{Numerical Simulations}
\label{sec:num_sims}
To facilitate the analytical analysis above, we have made numerous simplifying assumptions that we do not expect to be generally valid. However, it is of interest to test some of the concepts against hypothetical metasurface antennas just to get an sense of how predictive these naive models are. To test the predictions in the previous sections, we perform a series of numerical studies of the amplitude hologram apertures using a commercial, full-wave simulation software (CST Microwave Studio \cite{studio20143d}), which is based on the a finite integration technique (FIT). For these numerical studies, the same parameters as in Table \ref{tab:parameters_amp_only_beam} are used with the beam direction selected to be $\phi_0$=-20\degree. We consider here two types of holograms: binary amplitude and amplitude only.

\subsection{Binary Hologram Simulations}

In this section we consider a metasurface antenna that reproduces the behavior of the binary hologram. Although the metasurface hologram in Fig. \ref{fig:metasurface-antenna} is depicted with an array of cELC elements, the analytical theory is not limited to this particular type of metamaterial element. To demonstrate the applicability of the theory presented in this work for different metamaterial types, in addition to cELCs, we also study slot-shaped sub-wavelength metamaterial irises. Similar to the cELCs, the slots couple to the magnetic field of the reference wave and can be modeled as magnetic dipoles, each with a magnetic moment proportional to the magnetic field of the reference wave by the polarizability (Eq. \ref{eq:polarizability}). In Fig. \ref{fig:Binary_Design}, we present a waveguide-fed metasurface antenna implementation, using slots as the metamaterial elements desgined to produce a binary amplitude hologram.        

\begin{figure}[htbp]
\centering
\mbox{\includegraphics[width=\linewidth]{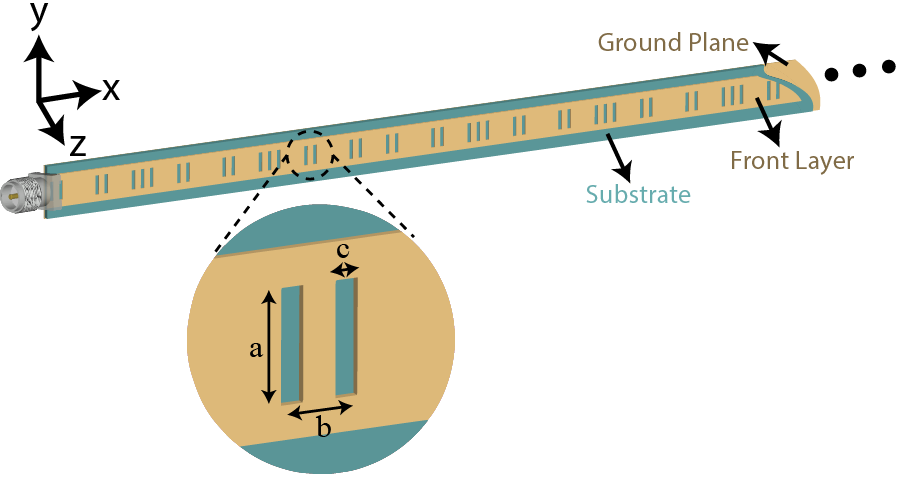}}
\caption{Designed binary amplitude hologram metasurface antenna with slot-shaped unit cells. To improve the clarity of the figure, only half of the antenna is shown here. Unit cell parameters are as follows: $a$=5.8 mm, $b$=3 mm and $c$=1 mm.}
\label{fig:Binary_Design}
\end{figure}

As shown in Fig. \ref{fig:Binary_Design}, the binary amplitude hologram design consists of a microstrip transmission line with the front surface of the aperture patterned with an array of subwavelength sized, slot-shaped elements. Each element can be modeled as a magnetic dipole along the longitudinal (y-) axis of the slot. As a result, as depicted in Fig. \ref{fig:metasurface-antenna}(b), the slots couple to the y-polarized magnetic field of the reference wave. The weight factors for the binary amplitude hologram are calculated using the same formulation as in Eq. \ref{eq:plane_wave_weights_amp_bin}, where for the positions corresponding to ``on" state we place a slot (coupling to the guided mode) while for the positions corresponding to "off" state, no slot is placed (no coupling). In a dynamically reconfigurable metasurface antenna, the slots could be dynamically switched between ``on" and ``off" states using a variety of tuning approaches, including active or nonlinear elements such as diodes, varactors, or transistors. To ensure that the slots exhibit weak coupling to the guided mode, such that the reference wave is not perturbed strongly, the lengths of the slots considered in this section do not exceed half of the guided-mode wavelength, $\lambda_g$/2. To be consistent with the analytical studies, as given in Table \ref{tab:parameters_amp_only_beam}, the waveguide index is selected to be 2.5, corresponding to $\epsilon_r$=6.25 for the dielectric substrate (non-magnetic). The thickness of the microstrip transmission line is 2 mm or $\lambda_g$/6, satisfying $<\lambda_g$/2 to ensure single mode operation while the characteristic impedance of the transmission line is chosen to be \textit{Z}=50$\Omega$ to match the feeding port impedance. The simulated radiation pattern of the binary amplitude hologram aperture is presented in Fig. \ref{fig:Binary_field_pattern}. The HPBW of the simulated binary hologram metasurface antenna is 3.73\degree while the first sidelobe level is -12.71 dB with the beam pointing at $\phi_0$=-20\degree, exhibiting good agreement with the analytical result presented in Fig. \ref{fig:beam_steer_bin}. The slight discrepancy between the analytical and simulated sidelobe patterns can be attributed to the weak perturbation of the phase of the reference wave due to the scattering from and coupling to the slot elements, which is not taken into account in the analytical model. The directivity of the simulated antenna is reported to be 13.9 dB.     

\begin{figure}[htbp]
\centering
\mbox{\includegraphics[width=\linewidth]{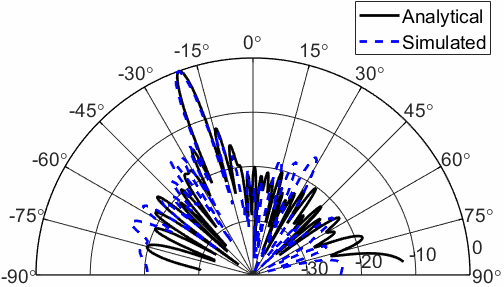}}
\caption{Simulated radiation pattern of the binary hologram metasurface antenna (dashed line) plotted on top of the analytical result of Fig. \ref{fig:beam_steer_bin} (solid line). Logarithmic scale (dB).}
\label{fig:Binary_field_pattern}
\end{figure}

Following the numerical analysis of the binary hologram metasurface antenna with slot-shaped metamaterial elements, we design the same binary hologram antenna but with cELC elements, as depicted in Fig. \ref{fig:cELC_design}. Similar to the metasurface antenna of Fig. \ref{fig:Binary_Design}, the same parameters as in Table \ref{tab:parameters_amp_only_beam} are used with the beam direction selected to be $\phi_0$=-20\degree.

\begin{figure}[htbp]
\centering
\mbox{\includegraphics[width=\linewidth]{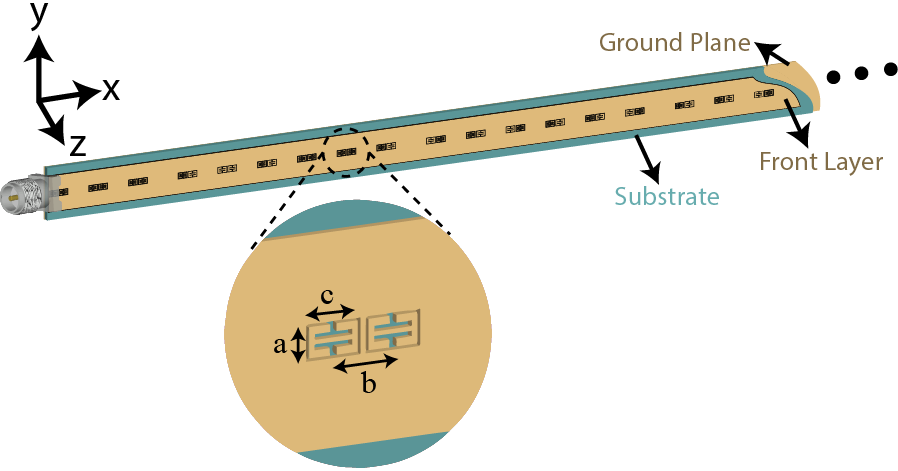}}
\caption{Designed cELC metasurface antenna using the binary amplitude hologram concept. To improve the clarity of the figure, only half of the antenna is shown here. Unit cell parameters are as follows: $a$=2 mm, $b$=3 mm and $c$=1.3 mm.}
\label{fig:cELC_design}
\end{figure}

The simulated radiation pattern of the binary amplitude hologram cELC metasurface antenna is shown in Fig. \ref{fig:cELC_field_pattern}.

\begin{figure}[htbp]
\centering
\mbox{\includegraphics[width=\linewidth]{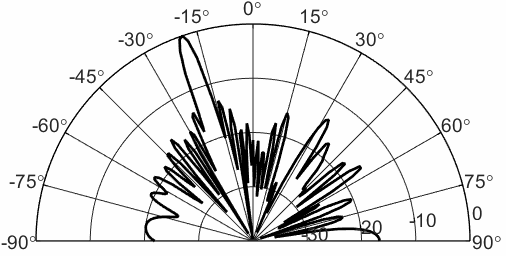}}
\caption{Simulated radiation pattern of the amplitude only hologram cELC metasurface antenna. Logarithmic scale (dB).}
\label{fig:cELC_field_pattern}
\end{figure}

Analyzing Fig. \ref{fig:cELC_field_pattern}, the HPBW of the simulated cELC binary hologram metasurface antenna is 3.7\degree while the first sidelobe level is -13.1 dB with the beam direction of the antenna being equal to $\phi_0$=-20\degree, as predicted from the analytical result presented in Fig. \ref{fig:beam_steer_bin}. Comparing the overall sidelobe levels in Fig. \ref{fig:beam_steer_bin} and Fig. \ref{fig:cELC_field_pattern}, it can be concluded that the numerical result of Fig. \ref{fig:cELC_field_pattern} exhibits slightly higher sidelobes. Similar to the metasurface antenna with slots, this discrepancy can be attributed to the perturbation of reference wave phase due to the interaction with the cELCs. It should be noted that the cELCs exhibit Lorentzian resonance response and can significantly alter the phase of the reference wave. Although this is a significant advantage for a phase hologram---as explained earlier---for an amplitude hologram it is desired that the phase of the reference wave is preserved, which is the case for the analytical result presented in Fig. \ref{fig:beam_steer_bin}. This distortion can be minimized by exciting the cELCs at a frequency close to the their resonance frequencies (not exactly at resonance), ensuring that the cELCs act as weakly coupled elements. Moreover, in comparison to the slot-shaped elements, which conventionally exhibit a wide resonance bandwidth \cite{balanis2016antenna,yurduseven_ieee_access_2017}, the cELCs have narrow-band resonance characteristics. When placed in an aperture as shown in Fig. \ref{fig:cELC_design} at close proximity from each other (with respect to the wavelength), they can strongly couple to each other (mutual coupling). The strong coupling of adjacent cELCs can shift their resonances, which, in return, can easily result in the cELCs resonating at undesired frequencies (rather than the intended frequencies), making them opaque patches at the desired operating frequency. Therefore, the design of the cELCs in the metasurface antenna of Fig. \ref{fig:cELC_design} requires significant attention to such details. 

\subsection{Amplitude-Only Hologram Simulations}

We next present a numerical study of the amplitude only hologram. For this study, we use the slot-shaped elements introduced above. Similar to the binary amplitude hologram study, we use the same parameters as in Table \ref{tab:parameters_amp_only_beam} and a beam direction of $\phi_0$=-20\degree. The weight distribution of the slots for this study is calculated using Eq. \ref{eq:plane_wave_weights_amp} with the weights ranging from 0.5 to 1.5. Different from the binary amplitude hologram where only ``on" and ``off" states are present, the amplitude only hologram exhibits a continuous variation of tuning states with varying coupling strengths governed by the weight distribution of the elements. We also refer to this amplitude distribution as a \textit{grayscale} amplitude topology. The coupling response, and therefore the weight, of a slot can be controlled by varying its length, which governs the resonance frequency. To understand the relationship between the slot geometry and its response, we first design a microstrip transmission line consisting of a single slot placed in the center, as shown in Fig. \ref{fig:single_unit_cell}. For this analysis, the length of the microstrip transmission line is selected to be 2$\lambda_g$ at 10 GHz.    

\begin{figure}[htbp]
\centering
\mbox{\includegraphics[width=\linewidth]{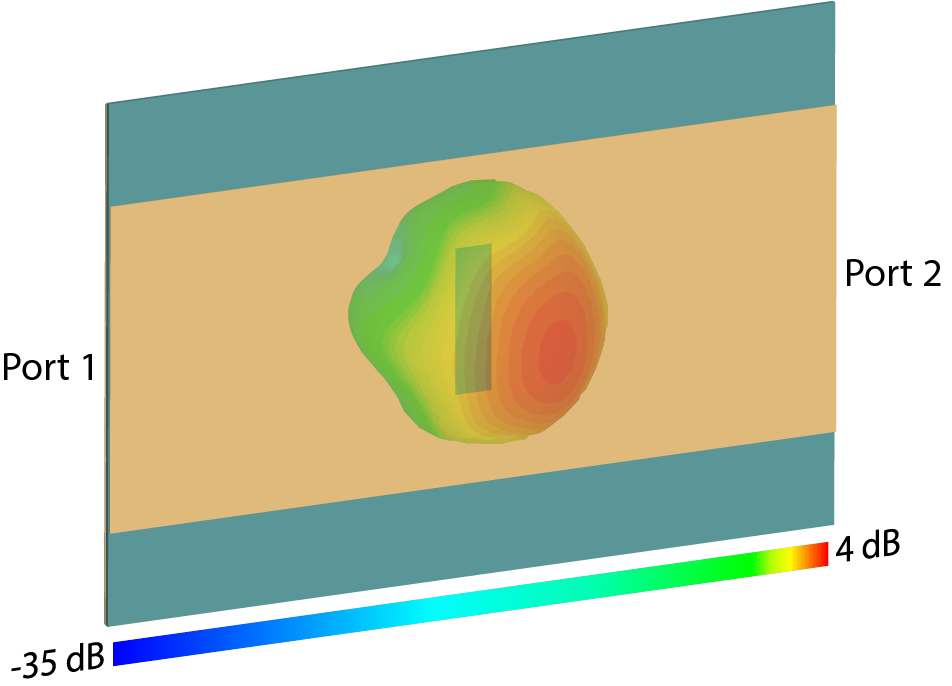}}
\caption{Simulation of a single unit cell. The field pattern radiated from the unit cell is overlaid on top.}
\label{fig:single_unit_cell}
\end{figure}

By performing the full-wave simulation of this structure, we analyze the amount of radiated power from the slot as a function of its length. For this simulation, we vary the length of the slot from $\lambda_g$/2 to $\lambda_g$/4 at 0.0125$\lambda_g$ intervals and measure the radiated power level as shown in Fig. \ref{fig:Slot_Length_vs_Power} and Table \ref{tab:radiated power levels}. It should be noted that although the parametric sweep was done at 21 intervals, in Table \ref{tab:radiated power levels}, we demonstrate only three cases (first, $\lambda_g$/2, center, $\lambda_g$/2.67, and last, $\lambda_g$/4) for clarity. For this analysis, the width of the slot is $\lambda_g$/10 while the total simulated power is 0.5 W.   

\begin{figure}[htbp]
\centering
\mbox{\includegraphics[width=\linewidth]{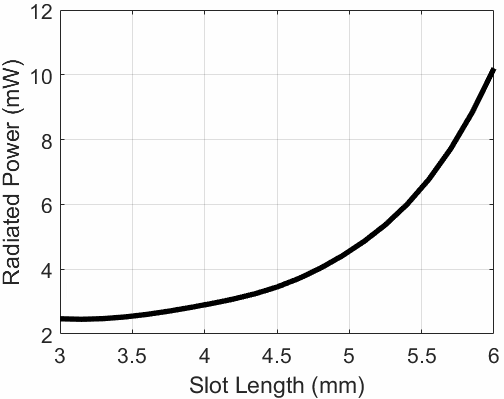}}
\caption{Quasi-quadratic behavior of the radiated power from the slots as a function of the slot length.}
\label{fig:Slot_Length_vs_Power}
\end{figure}

\begin{table}[htbp]
\centering
\caption{\bf Slot Radiated Power as a Function of Length}
\begin{tabular}{ccc}
\hline
Length & Radiated Power \\
\hline
6 mm ($\lambda_g$/2) & 10.4 mW \\
4.5 mm ($\lambda_g$/2.67) & 3.6 mW \\
3 mm ($\lambda_g$/4) & 2.5 mW \\
\hline
\end{tabular}
  \label{tab:radiated power levels}
\end{table}

Analyzing Table \ref{tab:radiated power levels}, it can be seen that the ratio between the radiated power levels for different slot lengths is approximately proportional to the square of the ratio between the corresponding slot lengths (quasi-quadratic behavior). For example, as shown in Table \ref{tab:radiated power levels}, reducing the slot length by a factor of 2 (from $\lambda_g$/2 to $\lambda_g$/4) reduces the radiated power level by a factor of 10.4/2.5=4.16. As the slot length approaches to $\lambda_g$/4, the power curve starts diverging from the ideal quadratic behavior due to the extremely weakened coupling of the slots to the guided mode reference. Since we are interested in the field coupling response of the elements, and the power is proportional to the product of the electric and magnetic fields, we observe the square-root difference between the radiated power levels (for this example, $\sqrt{4.16}$=2.04), which is approximately equal to the factor by which the slot length is changed (for this example, 2). As a result, we adjust the lengths of the slots with respect to the weight factors of Eq. \ref{eq:plane_wave_fp_amp} varying from 0.5 to 1.5---that is, the ratio between the longest and shortest slots in the hologram is 1.5/0.5=3. The designed amplitude-only hologram is demonstrated in Fig. \ref{fig:Grayscale_Design}. 

\begin{figure}[htbp]
\centering
\mbox{\includegraphics[width=\linewidth]{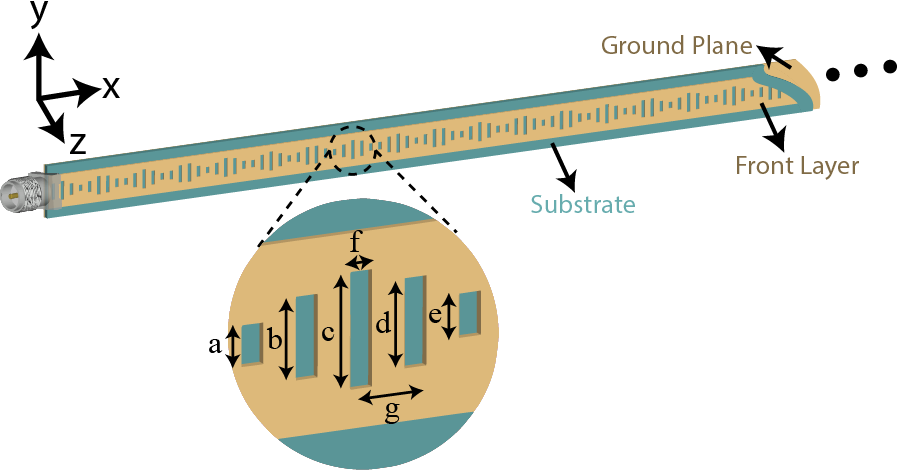}}
\caption{Designed metasurface antenna using the amplitude only hologram concept. To improve the clarity of the figure, only half of the antenna is depicted here. Unit cell parameters are as follows: $a$=2 mm, $b$=3.2 mm, $c$=5.9 mm, $d$=3.7 mm, $e$=2 mm, $f$=1 mm and $g$=3 mm.}
\label{fig:Grayscale_Design}
\end{figure}

The simulated radiation pattern of the amplitude-only hologram is shown in Fig. \ref{fig:Grayscale_field_pattern}.

\begin{figure}[htbp]
\centering
\mbox{\includegraphics[width=\linewidth]{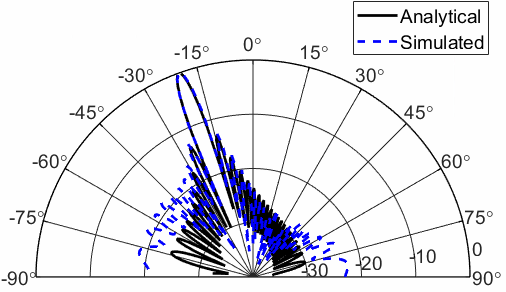}}
\caption{Simulated radiation pattern of the amplitude only hologram metasurface antenna (dashed line) plotted on top of the analytical result of Fig. \ref{fig:beam_steer_amp_only} (solid line). Logarithmic scale (dB).}
\label{fig:Grayscale_field_pattern}
\end{figure}

As can be seen in Fig \ref{fig:Grayscale_field_pattern}, the simulated hologram design produces a well defined beam pointing in the predicted direction, $\phi_0$=-20\degree. In comparison to the simulated radiation pattern of the binary amplitude hologram design shown in Fig. \ref{fig:Binary_field_pattern}, the overall sidelobe levels are lower and the pattern exhibits superior fidelity. The HPBW of the simulated amplitude only hologram metasurface antenna is 3.68\degree while the first sidelobe level is -12.86 dB, exhibiting good agreement with the analytical result presented in Fig. \ref{fig:beam_steer_amp_only}, albeit exhibiting slightly larger overall sidelobes, especially in the range between -90\degree and 0\degree. Similar to the numerical analyses of Figs. \ref{fig:Binary_field_pattern} and \ref{fig:cELC_field_pattern}, this can be attributed to the fact that, although weakly coupled, the full-wave numerical model of the antenna includes the distortion caused by the coupling of the elements to the reference wave, resulting in the phase of the reference wave diverging from the ideal analytical model which does not include this perturbation. Moreover, whereas the radiation pattern of the magnetic dipoles in the analytical model can be considered omnidirectional in the x-z plane, the simulated radiation pattern of the slot elements shown in Fig. \ref{fig:single_unit_cell} does not exhibit the ideal omnidirectional behavior due to the finite size of the ground plane, contributing to the slight discrepancy in the overall sidelobe levels, especially below -20 dB. The directivity of the simulated antenna is reported to be 14.5 dB, slighly larger than the directivity of the binary hologram metasurface antenna, 13.9 dB.         

\section{Conclusion}

We have presented an overview of the waveguide-fed metasurface antenna and provided a set of closed-form, analytical expressions that describe the essential function and radiative properties of the antenna. While numerous simplifying assumptions were made to facilitate the analysis, such as (a) weak scattering from the magnetic dipoles, (b) weak perturbation of the reference wave by the magnetic dipoles, and (c) no strong coupling between the magnetic dipoles, full-wave simulations on metasurface antennas display good agreement with the theory, predicting key antenna metrics, such as beam steering direction, HPBW, sidelobes, pointing accuracy, and other characteristics of interest. It should be noted that the antennas simulated were designed with the assumptions of the theory in mind, and that many other implementations may not produce results in such close agreement with theory. In such cases, it may be possible to extend the analytical framework by considering the coupling of the dipole elements through the waveguide modes and through the radiated fields. Such an approach has been used in the development of a modeling tool for slotted-waveguide leaky-wave antennas \cite{pulido_awpl_2016}, and can easily be extended to much larger apertures.

In the simulations presented here, no attempt was made to numerically extract the effective polarizability of a metamaterial element. Polarizability extraction would be a logical next step in the analytical modeling of the metasurface antenna, as it provides the exact value for a metasurface element and can easily be implemented to characterize experimental samples.

The waveguide-fed metasurface antenna provides considerable design flexibility that can be advantageous in many scenarios. Like a leaky-wave or traveling wave antenna, the metasurface antenna leverages the phase advance of the waveguide mode, avoiding the need for phase-shifting circuits that can add cost and complexity to the system. The absence of complete control over phase can be compensated, at least partially, by sampling the aperture as finely as is feasible, enabling a holographic design methodology to be pursued. We have not considered many key implementation questions, such as bandwidth, matching and many other details that will be of ultimate interest in applications. While some of these questions can indeed be addressed in part by the theory developed here, the range of possible systems and usage scenarios would make such an analysis more specific and hence beyond the scope of this analysis. Such details will be taken up in future work. The presented work exhibits a useful way of demonstrating how metasurface antennas perform beam forming and predicting their radiation characteristics. It can be considered as a simple, yet a compelling guideline to understand this promising concept and its significant potential for dynamic beam forming, opening up a host of new opportunities in applications ranging from satellite communications to wireless power transfer and radar imaging.

\section{ACKNOWLEDGMENT}
This work was supported by the Air Force Office of Scientific Research (AFOSR, Grant No. FA9550-12-1-0491). The authors also wish to acknowledge Casey Tegreene and Russell Hannigan (Invention Science Fund, Intellectual Ventures) for their important input and support.

% Bibliography
\bibliography{sample}

\end{document}